# Visualizing a dilute vortex liquid to solid phase transition in a $Bi_2Sr_2CaCu_2O_8$ single crystal

**Gorky Shaw[1], Pabitra Mandal[1], S. S. Banerjee[1,*] and T. Tamegai[2]**

[1]Department of Physics, Indian Institute of Technology, Kanpur-208016, India

[2] Department of Applied Physics, The University of Tokyo, Hongo, Bunkyo-ku, Tokyo 113-8656, Japan.

*E-mail: satyajit@iitk.ac.in

**Abstract:** Using high sensitivity magneto-optical imaging we find evidence for a jump in local vortex density associated with a vortex liquid to solid phase transition just above the lower critical field in a single crystal of $Bi_2Sr_2CaCu_2O_8$. We find the regions of the sample where the jump in vortex density occurs are associated with low screening currents. In the field – temperature vortex phase diagram we identify phase boundaries demarcating a dilute vortex liquid phase and the vortex solid phase. The phase diagram also identifies a coexistence regime of the dilute vortex liquid and solid phases and shows the effect of pinning on the vortex liquid to solid phase transition line. We find the phase boundary lines can be fitted to the theoretically predicted expression for the low-field portion of the phase boundary delineating a dilute vortex solid from a vortex liquid phase. We show that the same theoretical fit can be used to describe the pinning dependence of the low-field phase boundary lines provided a dependence of the Lindemann number on pinning strength is considered.

# 1. Introduction

The conventional [1] field ($H$) – temperature ($T$) phase diagram for an ideal type-II superconductor consists of a mixed state with a hexagonally ordered vortex lattice present between the lower critical field ($H_{c1}(T)$) and the upper critical field ($H_{c2}(T)$). Numerous studies have established that long range spatial order in the ideal hexagonal Abrikosov lattice is destroyed due to pinning in a non-ideal superconductor. The vortex state in a superconductor could either be a reasonably well ordered glassy phase devoid of topological defects like dislocations, viz., a Bragg glass, or be a disordered phase with topological defects, viz., the vortex glass phase [2,3]. At high $H$ when the inter-vortex spacing, $a_0 \propto \sqrt{\phi_0/B}$ (where $\phi_0$ is the quantum of magnetic flux = 2.07 x $10^{-7}$ G-cm$^2$ and $B \sim \mu_0 H$) is < the superconducting penetration length $\lambda$ (= range of vortex-vortex interaction), non-local effects lead to a softening of the elastic modulii of the vortex lattice [4,5] thereby making the vortex lattice susceptible to thermal fluctuation and pinning effects [2,5]. In high-$T_c$ superconductors, high superconducting transition temperatures ($T_c(0)$), high anisotropy and small condensation energy, result in the vortices being highly susceptible to thermal fluctuations. Across the melting phase boundary at high fields large thermal fluctuations cause the soft vortex solid (VS) to melt into a vortex liquid (VL) phase, which is characterized by zero shear elastic modulii [2,6,7]. Akin to ice to water transformation, the vortex solid to liquid melting transition produces an enhancement in the density of vortices. The melting at higher fields has been well studied both theoretically and experimentally. Experiments on melting at high $H$ have established the presence of a jump in the equilibrium magnetization [8] and the presence of latent heat [9] associated with the first order vortex solid to liquid melting phase transition. However, the nature of the vortex phase at low fields is less well investigated. At low fields (near $H_{c1}$), where $a_0 > \lambda$, the weak inter-vortex interaction leads to exponentially small elastic modulii [2,4] of the lattice making the vortex state susceptible to thermal fluctuations and pinning effects. In the $H$-$T$ phase diagram at intermediate fields one encounters the VS phase. It has been Theoretically proposed [2,6,10] that at fixed $T$ by increasing $H$ or decreasing $H$, the VS undergoes a phase transformation into the VL phase across a phase boundary close to $H_{c2}(T)$ and another phase boundary close to the $H_{c1}(T)$. The vortex solid therefore can melt at both high and low fields and in the $H$-$T$ vortex phase diagram the vortex solid phase is bounded by two lines, across which the vortex solid melts into a liquid phase. In the $H$-$T$ vortex phase diagram, at high fields, the line across which the vortex solid melts follows a monotonic behavior with negative slope tracking in the behaviour of $H_{c2}(T)$ albeit with a different temperature dependence. In the low-field regime near $H_{c1}(T)$ the line across which the dilute vortex solid transforms into a vortex liquid is associated with exponentially small shear modulus at low $H$ (For a schematic of the melting phase boundary see ref. [11]). Apart from thermal fluctuations at low magnetic fields due to the weakened shear modulus, the elastic vortex lattice is highly susceptible to the disordering influence of pinning. Experiments on low-$T_c$ superconductors with weak pinning have suggested a disordered vortex phase [12,13,14,15] present at low $H$ which is distinct from the elastic, ordered vortex phase present at higher fields. Recently scanning hall probe imaging of the vortex state in disordered thin films of BSCCO [16] at low fields (above 2.4 Oe) have captured of images with fluctuating contrast. The authors ascribe the fluctuating contrast to the presence of a vortex liquid phase, wherein the rapidly fluctuating vortices in the liquid phase are intermittently trapped and pinned on the strong pinning centers leading to enhancement in contrast and its subsequent decay as the vortex escapes from the pin. Using sensitive magneto-optical imaging technique we have identified a change in the equilibrium vortex density associated with a vortex liquid to solid phase transition at low fields near $H_{c1}$. We construct an $H$-$T$ vortex phase diagram identifying a line across which the vortex liquid transforms into a vortex solid phase via an intermediate regime of phase coexistence comprising of vortex liquid and solid phases. The low-field phase transformation boundary we have identified fits to the theoretically proposed expression [10] for the low-field line across which a vortex liquid transforms into a vortex solid phase. Using the magneto-optical imaging technique we have also spatially resolved the magnetization relaxation in different regions of the sample to enable the construction of a coarse map of the pinning landscape in the sample. We investigate the effect of pinning on the low-field vortex liquid to solid transformation line by showing a significant increase in the Lindemann number with pinning strength.

## 2. Experimental details

We chose a high quality single crystal of $Bi_2Sr_2CaCuO_8$ (BSCCO) [17] of dimensions (0.8 x 0.5 x 0.03 $mm^3$) and $T_c$ = 90 K. To introduce vortices in the sample, magnetic field was applied parallel to the *c*-axis of the single crystal ($H \| c$) using a copper coil solenoid magnet. High sensitivity and high field homogeneity across the sample is maintained within the solenoid in the field range 0 - 200 Oe. We employ the conventional magneto-optical imaging (MOI) [18] as well as the differential magneto-optical (DMO) imaging technique [19,20] to image the changes in local vortex density. A schematic and details of our MOI setup have been presented elsewhere [21]. The DMO technique is sensitive to imaging small changes ($\delta B_z(x, y)$) in local field distribution $B_z(x, y)$ (where ($x, y$) defines the sample plane perpendicular to $H$, and $z$ is along $H$). We obtain differential ($\delta(x, y)$) images by increasing $H$ by an amount $\delta H$ = 1 Oe and capturing Faraday rotated images $I_i(H)$ and $I_i(H+\delta H)$, *i* number of times at $H$ and $H+\delta H$, $\delta(x, y) = \frac{\sum_{i=1}^{k} I_i(H+\delta H)}{k} - \frac{\sum_{i=1}^{k} I_i(H)}{k}$, where usually $k$ = 20. Appropriate calibration converts $\delta(x, y)$ to provide a measure of $\delta B_z(x, y)$ [19,20]. Unlike conventional DMO [19], due to irreversible magnetization response of BSCCO at low fields, we usually do not average $\delta(x, y)$ by repeated modulation of the external field by $\delta H$ about $H$. In the $\delta(x, y)$ images (for example see fig. 1 below), the bright and dark contrasts correspond to high and low $\delta B_z$.

## 3. Results and discussion

### *3.1. DMO and evidence for jump in $B_z$ at low H*

Figures 1(a) to 1(d) show $\delta(x, y)$ (DMO) images captured at 50 K with $H$ = 30 Oe, 36 Oe, 42 Oe and 57 Oe respectively. Across the (black) vertical line drawn on the images in figs. 1(a) – 1(d), the $\delta B_z(r)$ behaviour (*r*: distance along the line) is measured at different $H$ (at the same *T*), and plotted in fig. 1(e). Each $\delta B_z(r)$ plot (different colour for each $H$) in fig. 1(e) has been artificially offset by 3 G for the sake of clarity. The bright contrast observed along the sample edges (cf. fig. 1(a)) is due to strong edge screening currents (discussed later in the context of fig. 2(a) and fig. 4(b)). The gray contrast outside the sample and away from the sample edges in the differential images corresponds to the change in the Faraday rotated magneto-optical intensity due to the increase in $H$, viz., $\delta H$ by 1 Oe. Note that the $\delta B_z(r) \sim 1$ G far away from the sample edges (viz., close to the 0 μm tick in graph shown in fig. 1(e)). At 30 Oe (fig. 1(a)), due to strong diamagnetic screening of flux inside the superconductor, most of the regions inside the sample possess a dark contrast, where $\delta B_z << 1$ G. This corresponds to the vortex free diamagnetic Meissner state. Flux enters the sample interior through a cigar shaped arm viz., cf. gray location near * marked in fig. 1(a) at 30 Oe. As $H$ is increased by 1 Oe, viz., $\delta H$ = 1 Oe, the local $\delta B_z$ near * increases by $\sim 1$ G, as seen in fig. 1(e) (red curve). As $H$ is increased to 36 Oe, $\delta B_z$ abruptly increases to $\sim 2$ G (viz., see blue curve in fig. 1(e) and the corresponding bright region around * in fig. 1(b)). As $H$ is increased further to 42 Oe (green curve in fig. 1(e)), $\delta B_z$ near * region (fig. 1(c)) decreases back to a value $\sim 1$ G while the region with enhanced mean $<\delta B_z> \sim 2$ G spreads to other locations of the sample where flux has already penetrated (cf. double arrow headed vertical line marked on green curve in fig. 1(e)). At 57 Oe the enhanced $<\delta B_z>$ reduces down to 1 G = $\delta H$ = 1 Oe, which corresponds to the onset of a weakly pinned vortex phase (viz., the vortex solid (VS) phase), uniformly over the entire sample (cf. fig. 1(d) and the corresponding pink curve of fig. 1(e)). Figure 1(f) compares the $\delta B_z$ signal at 36 Oe and 57 Oe in a 50 μm wide region in the sample, viz., a region between 550 – 600 μm in the line scans of fig. 1(e), where the anomalous jump in $\delta B_z$ occurs at 36 Oe. It is clear from fig. 1(f) that $\delta B_z$ is anomalously enhanced at 36 Oe compared to that at 57 Oe. As will be shown later in fig. 2(b) that at 50 K, upon increasing the field from 36 Oe to 57 Oe, a VL phase transforms into a VS phase. At this juncture it is worth mentioning that observation of a brightening in a DMO image corresponding to an enhanced $\delta B_z$ over and above the background intensity, is used to signify [19,22] a vortex phase transformation associated with a change in the density of vortices. The ($H$, $T$) where $\delta B_z$ signal locally becomes maximum (and well above $\delta H$ = 1 Oe) is situated deep inside the vortex liquid phase and signals the transformation into the vortex solid phase. This behavior occurs reproducibly and repeatedly at the same location in the sample over repetition of the above measurements.

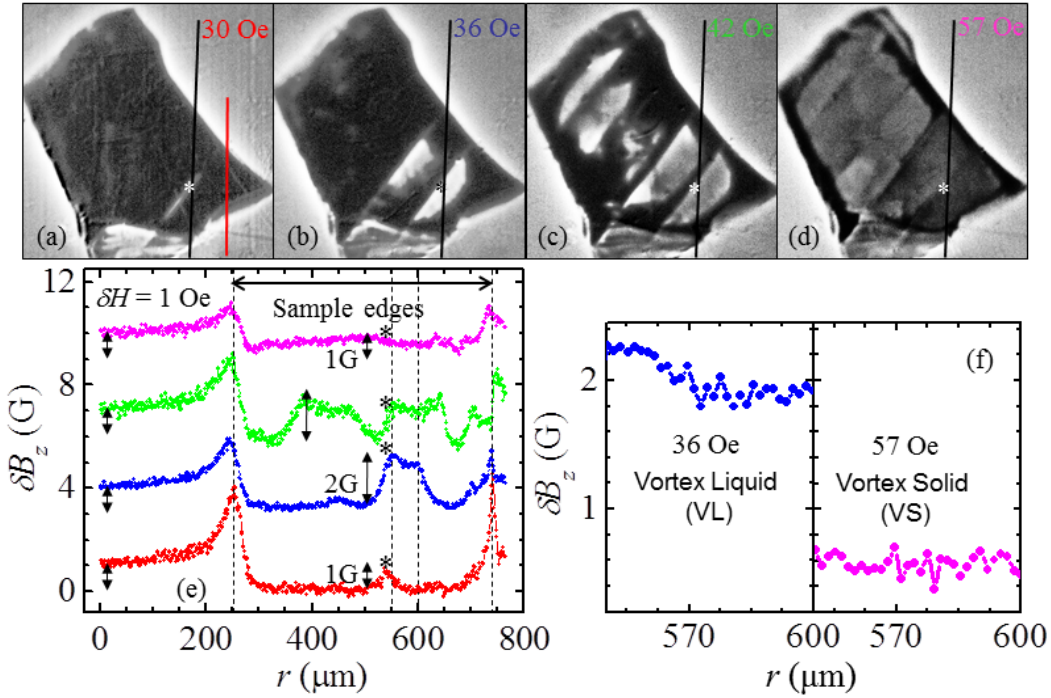

**Figure 1.** (a)–(d) Differential MOI ($\delta(x, y)$) images captured at 50 K with $H$ at 30, 36, 42 and 57 Oe respectively. (e) $\delta B_z$ vs. $r$ (: distance along the line) behaviour across the black line in the above images. The $\delta B_z(r)$ plots are artificially shifted by 3 G for the sake of clarity. The double headed arrows represent the size of $\delta B_z$ at different locations along the $\delta B_z(r)$ plots. (f) Comparison of $\delta B_z(r)$ at 36 Oe and 57 Oe in the region 550 - 600 μm (indicated by the respective dashed lines in (e)). The artificial shifts incorporated in fig. 1(e) for clarity have been removed while plotting the data in fig. 1(f).

*3.2. Nature of vortex distribution in the sample and determination of width of the jump in $B_z$*

Along with differential $\delta B_z(x, y)$ images we have also measured the $B_z(x, y)$ distribution across the sample using conventional MOI (images not shown). Fig. 2(a) shows $B_z(r)$ variation with increasing $H$ across the red line shown in fig. 1(a), with the distance ($r$) being measured along the red line. In a sample with weak pinning and large edge currents, vortices are pushed towards the center of the sample resulting in the dome shaped field profile [23]. From the $B_z(r)$ curves in fig. 2(a), we observe large $B_z$ near the sample edges indicative of barriers associated with large screening currents flowing at the sample edges (discussed later on in fig. 4(b)). Inside the sample we observe a dome shaped feature in $B_z(r)$ developing as the vortices locally penetrate into the sample as $H$ is increased. As $H$ increases from 28.5 Oe in steps of 1.5 Oe, initially the $B_z(r)$ curves remain flat due to a flux free Meissner phase. Beyond 31.5 Oe, the $B_z(r)$ curve develops a dome like feature (see the progressive development of the dome along the dashed line marked 1 near the $r$ = 200 μm tick). As expected, with every increment in $H$ by 1.5 Oe, the peak height of the dome in $B_z(r)$ increases. To estimate the amount of change in $B_z$ across different regions of the dome for a 1.5 Oe increment in $H$, we calculate the difference in the $B_z(r)$ value (viz., $\delta B_z$) between successive points located at the intersection of the vertical line 1 (or line 2) and the $B_z(r)$ curves in fig. 2(a), viz., $\delta B_z \vert_{1,2} = B_z \vert_{1,2} (H + \delta H) - B_z \vert_{1,2} (H)$ where $\delta H = 1.5$ Oe and $|_{1,2}$ represents points of intersection along line 1 or 2 in fig. 2(a). From fig. 2(b) we note that the maximum $\delta B_z|_{1,2}(H)$ is ~ 3 G with $\delta H = 1.5$ Oe is comparable to the $\delta B_z \sim 2$ G noted earlier in the DMO imaging procedure (fig. 1). In fig. 2(b) beyond $H_{c1} \sim 32$ Oe, $\delta B_z|_{1,2}$ begins deviating systematically from 0 G. As $H > H_{c1}$, for the curve labeled 1 in fig. 2(b) (black squares), $\delta B_z|_{1,2}$ increases and crosses a value of 1.5 G before reaching a maximum value of ~ 3 G at $H_M = 39$ Oe. Beyond 39 Oe $\delta B_z|_{1,2}$ begins to decrease from ~ 3 G and settles down to a constant value of ~ 1.5 G = $\delta H$. In fig. 2(b), the black dashed curve through the data points is a guide to the eye representing the above behaviour for the $\delta B_z$ found across line 1 in fig. 2(a), viz., $\delta B_z|_1$. At higher $H$ (45 Oe and above) the anomalous enhancement in the peak height of $B_z(r)$ ceases, cf. fig. 2(a), where above 45 Oe the consecutive $B_z(r)$ profiles (taken in intervals of 1.5 Oe) are almost equi-spaced along $B_z$ axis. At fields above 50 Oe in fig. 2(b) $\delta B_z|_{1,2} = 1.5$ Oe = $\delta H$, therefore in fig. 1(d) at 57 Oe, one observes a uniform grayish contrast acquired over the entire sample. In comparison to line 1, across line 2 (see fig. 2(a)), the peak in $\delta B_z|_{1,2}$ occurs later, viz., at $H_M = 40.5$ Oe (see red circles with the dotted red curve in fig. 2(b)). The $\delta B_z(r)$ behaviour across line 2 shows that behaviour similar to line 1 exists at different regions of the sample, and the peak in $\delta B_z(r)$ which corresponds to a brightening in the DMO image occurs at slightly different fields at different regions of the sample. The above discussion suggests that the jump in $\delta B_z$ progressively spreads to neighbouring regions of the sample with increasing $H$. The horizontal dashed line in fig. 2(b) represents $\delta B_z = \delta H$, viz., the line across which the amount of change in local flux density is equal to the change in external field. Note for the sake of comparison we have shown with a dotted blue curve in fig. 2(b), a representative sketch of a monotonic variation in $\delta B_z$ with increasing $H$ for depicting a hypothetical situation where no jump in $B_z$ exists, viz., the VS phase is gradually attained as the field is increased from $H_{c1}$. Note that both $\delta B_z|_{1,2}(H)$ curves in fig. 2(b) curves exceed the dotted blue curve. We would like to mention that by comparing figure 1 with the $B_z(r)$ profile in fig. 2 it is clear that the jump $\delta B_z \sim 2$ G noted in fig. 1 occurs in a region of the sample where vortices have already penetrated and the dome shaped $B_z(r)$ profile is established at a lower field. For example near the location * in the sample where the jump (bright contrast) in $\delta B_z$ has been noted at 36 Oe (fig. 1(a, b)), vortices had already penetrated into this region prior to 36 Oe as indicated by the presence of the dome shaped profile from 32 Oe onwards (cf. fig. 2(a)). At low fields just above $H_{c1}$, we propose that the change $\delta B_z \sim 2$ G observed in the DMO images is associated with the appearance of a first order like vortex phase transition from a VL to a VS phase. The peak attained in $\delta B_z|_{1,2}(H_M)$ signal in fig. 2(b) beyond $\delta B_z = \delta H$ indicates a phase which can sustain greater changes in vortex density compared to a VS phase, viz., we identify the peak $\delta B_z(H)$ signal corresponding to the limit of a low-field VL phase above $H_{c1}$ (we identify this limit as $H_M$ in fig. 2(b)). Beyond the maximum in $\delta B_z|_{1,2}$ the signal gradually decreases with increasing $H$ (> $H_M$), which corresponds to the regime of coexistence of a VL and a VS phase, before saturating to a value = 1.5 G = $\delta H$. In fig. 2(b) the field regime of VL and VS coexistence for line 1 is identified with a cross shaded region, while the coexistence field regime for line 2 overlaps with that of line 1 and is identified with a transparent gray shaded region overlapping the cross shaded region. Beyond the coexistence region we identify the VS phase in fig. 2(b). We would like to emphasize that that above 55 Oe (inside the VS phase) no further brightening in the DMO image is found in response to $\delta H$ modulation suggesting that the brightening observed in fig. 1 is due to an additional enhancement in $\delta B_z$ over and above the conventional effect of $B_z(r)$ dome height modulation when the external $H$ is modulated.

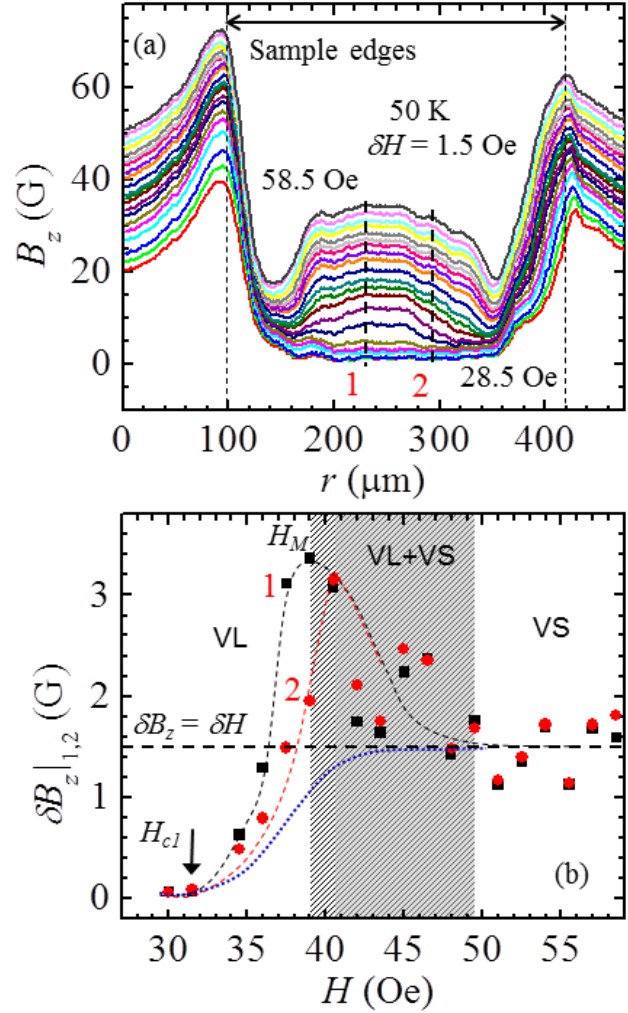

**Figure 2.** (a) Variation of $B_z$ along the red line marked in fig. 1(a) for $H$ being increased from 28.5 Oe to 58.5 Oe in steps of 1.5 Oe (b) The black squares and red circles represent variation of $\delta B_z|_{1,2}$ with $H$ across the two lines labeled 1 and 2 respectively in (a). The black and red dashed curves are guides to the eye. The notation $\delta B_z|_{1,2}$ represents $\delta B_z$ determined across line 1 or 2 in fig. 2(a). The horizontal dashed line represents the $\delta B_z = \delta H$ line (cf. text for details). The dotted blue curve is a schematic of the $\delta B_z$ vs. $H$ behaviour expected in a hypothetical situation where there exists no jump in $B_z$. The striped and shaded regions indicate an intermediate VL+VS coexistence regime over which the VL phase transforms into the VS phase, for lines 1 and 2 respectively (cf. text for details).

*3.3. Location of jump in $B_z$ at low H with respect to the vortex solid phase*

Figure 3(a) shows $M(H) = \dfrac{\int_A [B_z(x,y) - H] dx dy}{4\pi A}$, where $A$ is the area over which averaging is performed and the local field $B_z(x, y)$ is determined from conventional magneto-optical imaging (without DMO) [18,21]. $M(H)$ for whole sample corresponds to $A$ = entire sample area, while the local $M(H)$ for * corresponds to averaging over a region around the location marked with * in fig. 1 with area, $A = 25$ μm$^2$. From the initial linear portion of the $M(H)$ curves (blue and red curves for bulk and local responses respectively) it is evident that vortex penetration field is higher for the bulk. The higher bulk penetration field is associated with surface and geometric barriers [24,25,26], whose evidence was found in fig. 2(a) where one can identify large field gradients near the sample edges. The local $H_{c1}$ at 50 K is determined to be 30 Oe from the conventional deviation from a linear $M(H)$ behaviour (see red curve in fig. 3(a)), which is consistent with the value determined in fig. 2(b). In the local $M(H)$ (red) curve, after $H_{c1} \sim 30$ Oe, the local diamagnetic response shows an unconventional sharp drop. The sharp drop in diamagnetic magnetization curve in fig. 3(a) coincides with the jump in $\delta B_z$ noted in fig. 1, which as discussed above corresponds to an enhancement in the local vortex density ($\rho = \dfrac{B_z}{\phi_0}$). The jump in $\delta B_z$ suggests that in regions of the sample where vortices have penetrated, the local vortex density is anomalously enhanced as the external magnetic field is increased just above $H_{c1}$ indicating the presence of the VL phase. On further increasing $H$ beyond the sharp drop, the local (red curve) diamagnetic response begins to increase and tends to approach the bulk diamagnetic magnetization value (viz., the red curve shows a tendency to increase towards the blue curve in fig. 3(a)) with a bump like feature. Over this region in fig. 2(b) the $\delta B_z$ is $> \delta H$, though it gradually decreases towards $\delta H$ with increasing $H$. Around 60 Oe and beyond, the local (red) $M(H)$ curve monotonically decreases akin to the bulk (blue) $M(H)$ curve. The monotonic decrease in $M(H)$ with increase in $H$ occurs for a weakly pinned VS phase [27] due to weakening of the diamagnetic screening currents with increasing $H$. Recall that in the field regime of around 60 Oe and beyond, we have already noted the presence of a VS phase (identified with $\delta B_z = \delta H = 1.5$ Oe) in fig. 2(b). In DMO, the observation of $\delta B_z = \delta H$ which is characteristic of the onset of a reversible, weak pinning VS phase [20], appears in the field regime of around 60 Oe. The $B_z(r)$ line scan in fig. 2(a) shows the presence of a dome shaped field profile above 57 Oe at 50 K, which is characteristic of uniform vortex distribution associated with a weakly pinned VS phase in the interior of the sample. Furthermore, subsequently (figs. 4(a) and 4(b)) we also show the absence of significant bulk screening currents in the sample interior in the vicinity of 60 Oe. The sharp drop in $M(H)$ above $H_{c1}$ along with a the bump like feature in fig. 3(a) (red curve) is reminiscent of the Second Magnetization Peak anomaly observed in HTSC's [28]. It may be recalled that earlier suggestions [15] for the presence of a low-field disordered vortex state in low-$T_c$ superconductors had detected the presence of similar second magnetization peak like anomaly close to $H_{c1}$. Here, the bump like feature in the local $M(H)$ curve demarcates the approximate $H$ range over which the VL and VS phases coexist (identified with the cross-shaded region in fig. 3(a), (a similar shaded area in a different location of the sample has been shown in fig. 2(b)). The location of $H_M$ (cf. fig. 2(b)) is indicated by an arrow in fig. 3(a).

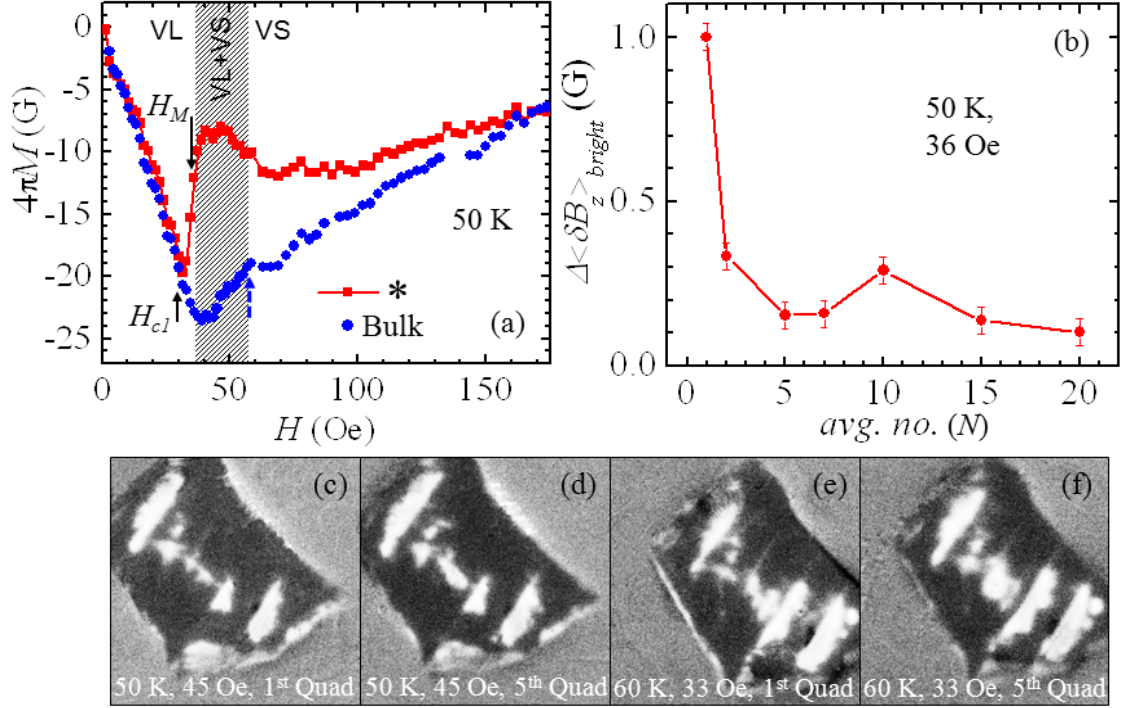

**Figure 3.** (a) $M(H)$ estimated from MOI images, locally in an area $\sim 25$ μm$^2$ around the * marked in 1(a)-1(d) (red squares), and for the whole sample area (blue circles) (cf. text for details). The shaded region indicates the approximate VL+VS coexistence regime corresponding to the local $M(H)$ curve. $H_M$ is the $H$ at which the maximum jump in $\delta B_z$ is found in the corresponding DMO images. (b) Variation of $\Delta \langle \delta B_z \rangle_{bright}$ vs. $N$, the no. of averaging done to obtain the DMO images, at 36 Oe, 50 K (cf. text for definitions and details). (c) – (f) DMO images obtained from 5-quadrant measurements (cf. text for details).

Due to the proximity of $H_M(T)$ to $H_{c1}(T)$ in fig. 3(a), one may suspect the change in $\delta B_z$ to be related to the flux change near $H_{c1}(T)$. Shown in the panel of images below fig. 3(a) are DMO images at the same $T$ for 50 K and 60 K in the first (I) and fifth (V) quadrants of the isothermal magnetization hysteresis loop where $H$ is varied over five quadrants. Clearly, the anomalous brightening in $\delta B_z$ observed at 45 Oe (50 K) in the first (viz., the virgin run) quadrant DMO image in fig. 3(c), which corresponds to a regime just above $H_{c1}(T)$ (cf. fig. 3(a)), is reproduced almost identically at $H$ = 45 Oe (50 K) in fig. 3(d) where the measurement is performed in the fifth quadrant (viz., measurements performed while increasing the $H$ from 0 G to the desired $+H$ value, where prior to increasing $H$ from 0 G, the field was originally cycled across four quadrants viz., from 0 G to +200 G (I), +200 G to 0 G (II), 0 G to -200 G (III) and -200 G to +0 G (IV)). The same is confirmed by the observation of similar anomalous brightening in measurements performed at different temperatures, as seen in the DMO images in figures 3(e) and 3(f) which are obtained at 60 K, 33 Oe in the first and fifth quadrants respectively. Therefore the brightening associated with the enhancement in $\delta B_z$ at $H_M(T)$ is not related to $H_{c1}(T)$. Note that in the second quadrant, due to strong irreversible magnetization response which leads to a positive magnetization background, it becomes difficult to discern the jump in magnetization when reducing the magnetic field from the VS phase in the low $H$ regime.

*3.4. Estimating the equilibrium height of jump in $B_z$*

While the jump in $B_z$ provides evidence for the existence of a VL phase prior to the onset of a VS phase, it is possible that the jumps measured in $\delta B_z$ in fig. 1 or 2 need not be the equilibrium value associated with this transition due to the metastable effects associated with pinning. To estimate the size of equilibrium jump in magnetization corresponding to a dense VL phase transforming into the VS phase (cf. fig. 1(f)), in fig. 3(b) we investigate the effect on $\delta B_z$ of repeated modulating by $\delta H$ = 1 Oe about $H$ ($N$ times), viz., we obtain the average differential image, $\langle \delta(x,y) \rangle_N = \dfrac{\sum_{i=1}^{N} \delta_i(x,y)}{N}$ at a given $H$ and $T$. From $\langle \delta(x,y) \rangle_N$ we define $\Delta \langle \delta B_z \rangle_{bright}$ = $\delta B_z$ in a region in the $\langle \delta(x,y) \rangle_N$ image with bright contrast (like the * region in fig. 1(b)) – 1 G (where 1 G is the $\delta B_z$ produced in response to $\delta H$ = 1 Oe in a reversible $H$, $T$ regime). In fig. 3(b) for $H$ = 36 Oe and $T$ = 50 K we plot $\Delta \langle \delta B_z \rangle_{bright}$ vs. $N$ in the * region where one observes a bright contrast (cf. fig. 1(b)). From fig. 3(b) we see beyond $N$ = 10, the $\Delta \langle \delta B_z \rangle_{bright}$ approaches an asymptotic non-zero constant value of about 0.1 G, which implies over and above the 1 G change in $\delta B_z$ due to $\delta H$ = 1 Oe, there exists and additional equilibrium enhancement in $\delta B_z$ ~ 0.1 G. In a region of the sample where a uniform vortex solid phase has formed, for example at 57 Oe, at 50 K in fig. 1(d), over most of the sample, $\Delta \langle \delta B_z \rangle_{bright}$ = 0. The quantity $\Delta \langle \delta B_z \rangle_{bright}$ is a measure of the jump in equilibrium vortex density due to a denser VL phase which is different from the VS phase that sets in at higher fields. Inset (b) of fig. 5 shows a DMO image from an isofield measurement, where we keep $H$ fixed and vary $T$. At each $T$ we modulate the temperature between $T$ and $T+\delta T$, where $\delta T$ = 0.4 K, and obtain differential MO images. We observe bright regions in the sample at 50 K in this varying temperature measurement in a constant dc field of 42 Oe which coincides with similar bright regions observed at 42 Oe in the varying magnetic field experiment at 50 K in fig. 1(c) (also reproduced as fig. 5(c)), indicating that the brightening feature observed is not path dependent in the $H$-$T$ space, but is a signature of a thermodynamic phase transformation.

*3.5. Distribution of screening currents in the low-field regime*

Using an inversion technique [29,30] we determine the distribution of the absolute value of the superconducting screening current density, $j(x,y)$ = $|\mathbf{J}(x,y)|$ = $|\sqrt{j_x(x,y)^2 + j_y(x,y)^2}|$, across the sample from $B_z(x, y)$ obtained from the conventional MO images, where $\mathbf{J} = j_x(x,y)\hat{x} + j_y(x,y)\hat{y}$ is the net current density at $(x, y)$ and $j_x$ and $j_y$ are the $x$ and $y$ components of the current density. Image in fig. 4(a) shows the $j(x, y)$ distribution across the sample at 57 Oe and 50 K (cf. fig. 1(d)).

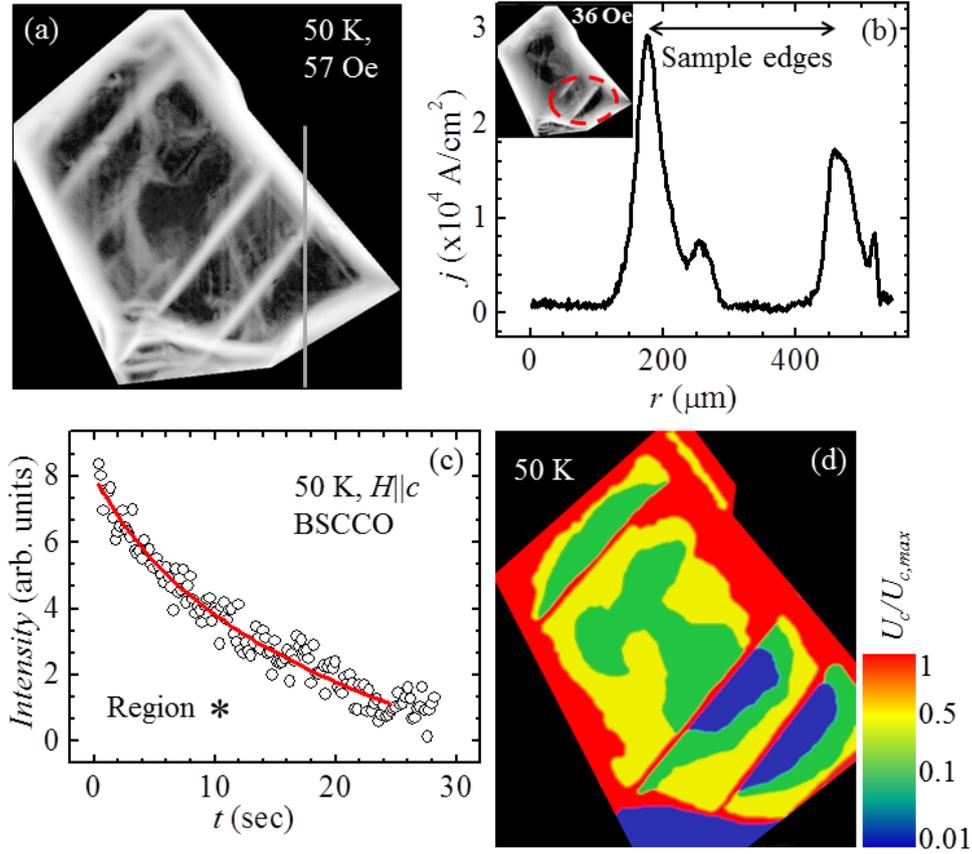

**Figure 4.** (a) Image showing the $j(x, y)$ distribution across the sample at 50 K, 57 Oe. (b) Variation of $j$ across the line shown in (a). Inset of (b) shows the $j(x, y)$ across the sample at 50 K, 36 Oe. (c) MOI intensity counts vs. $t$ at 50 K in the remnant state after field cooling in 150 Oe, averaged over a small region around the * marked in fig. 1(a). (d) The coarse pinning landscape $U_c(x, y)$ across the sample surface correlated with regions where the vortex liquid phase exists at different $H$. The adjoining colour bar indicates the normalized $U_c$ values (cf. text for details).

The bright regions in the image represent significant $j(x, y)$, while dark regions correspond to negligible $j(x, y)$. Note from the line scan in fig. 4(b) that the $j(x, y)$ across major portions of the sample is almost zero while large $j(x, y) \sim 10^4$ A/cm$^2$ is confined only along the sample edges and along the linear (diagonal) defects inside the sample which are regions with enhanced pinning or large screening currents. Such linear defects in single crystals of BSCCO are microscopic regions with compositional inhomogeneities incorporated into the crystal during its growth and have been reported earlier [22,31]. These regions possess enhanced vortex density compared to their neighboring regions [31] and as discussed later they may also act as nucleation centers for initiating the low-field melting of the vortex state. From the earlier discussion of figs. 2(b) and 3(a), we believe that $H$ around 60 Oe and beyond, (like 57 Oe in fig. 4(a)), corresponds to field values at which interactions between the vortices become sufficiently strong to promote the formation of a vortex solid phase. The onset of an equilibrium vortex distribution produces negligible gradients in the vortex distribution which is seen as low $j(x, y)$ as in fig. 4(a) present over majority of the sample interior where the vortex solid exists. Inset of fig. 4(b) shows an image of the $j(x, y)$ distribution at 36 Oe, at 50 K, in which we observe large diamagnetic screening currents distributed across the Meissner like regions of the sample. However it is noteworthy that the bright regions where a jump in $\delta B_z$ was observed in fig. 1(b), have low $j(x, y)$ values (cf. within the region encircled by the red dashed curve in fig. 4(b) inset). We reaffirm our original inference that the brightening observed in fig. 1 upper panels and figs. 3(c) – 3(f) is a feature which is not associated with penetrating vortices. Had the brightening in $\delta B_z$ been only related to penetrating vortices, then we should have observed large gradients and hence large screening currents $j(x, y)$ in regions where the jump occurs, which is absent as observed in fig. 4. The above is true for all the bright regions we have observed and for those shown in figs. 1 and 3.

*3.6. Mapping the pinning landscape and the location of regions with jumps in $B_z$*

By capturing images in interval of 200 ms, using conventional MOI technique we capture the time ($t$) evolution of $B_z(x, y, t)$ and hence of $M(x, y, t)$ of the remnant magnetized state of the superconductor (created from 150 Oe). Figure 4(c) shows the typical relaxation of the MOI intensity count vs. $t$ at 50 K, observed in the small region of area 25 μm$^2$ located in a region marked * in fig. 1(a). The intensity is calibrated to obtain the remnant magnetization $M_{rem}(x, y, t)$. Note that the magneto-optical intensity $I(x, y, t)$ is related to $M_{rem}(x, y, t)$ as $M_{rem}(x, y, t) \propto [I(x, y, t)]^{0.5}$. By fitting the $M_{rem}(x, y, t)$ data to the expression [32] $M(t) = M(0)(1+\frac{t}{t_0})^{-2K_BT/U_c}$ (see solid line through data in fig. 4(c)), we determine the local pinning potential $U_c$. Using the above procedure, from the relaxation of $M_{rem}$ determined at different positions in the sample, we obtain the coarse pinning potential landscape $U_c(x, y)$ in the sample. The coarse graining is over a 100 × 100 μm$^2$ area. The $U_c$ values are normalized to the maximum value ~ 85$k_B$ found in the sample and colour-coded as shown in fig. 4(d). Fig. 4(d) reveals that the enhanced $\Delta \langle \delta B_z \rangle_{bright} = 0.1$ G is primarily nucleated first along the linear defect regions with large $U_c$ (red). From here, it first spreads to the nearby regions with lower $U_c$ (cooler (blue) coloured regions), and then finally to other regions with larger $U_c$ (the warmer (red) coloured regions in fig. 4(d)). By comparing images of fig. 1 with the map of the pinning landscape in fig. 4(d), it appears that the vortices penetrate into the superconductor preferably through the regions with linear defects. Subsequently the jump in $\delta B_z$ is also nucleated in the same neighbourhood of linear defects after which the jump propagates through the weaker pinning regions of the sample before reaching the stronger pinning regions of the sample at higher fields.

The jump in $B_z$, viz., $\Delta \langle \delta B_z \rangle_{bright}$, which is the contrast difference encountered in the VL measured w.r.t. to the VS phase (with $\delta B_z = 1$ G), is associated with the abrupt enhancement in local vortex density which we propose is due to melting into a low-field vortex liquid phase from a higher field VS phase. Recall here that such jumps in local field have been observed for melting into the vortex liquid phase at high fields [8,19]. Furthermore, as mentioned earlier, $j(x, y)$ distribution shows that regions with $\Delta \langle \delta B_z \rangle_{bright} = 0.1$ G are devoid of screening currents, which is characteristic of a vortex liquid phase associated with almost zero critical current density due to the predominance of thermal fluctuations overcoming pinning in this phase. Figure 1 suggests that after the vortices enter the superconductor above $H_{c1}$ one encounters a thermal fluctuation dominated VL phase which at higher fields (viz., at higher vortex densities) transforms into a VS phase via a first order like transition. Referring to fig. 3(b), above $H_{c1}$ the density of the VL phase is higher than the VS phase at higher fields viz., it is $\rho + \delta\rho$, where this increase in vortex density, $\delta\rho$, associated with the equilibrium $\Delta \langle \delta B_z \rangle_{bright} \sim 0.1$ G jump, is due to the presence of a denser VL phase compared to the VS phase. With further increase in $H$, the excess $\delta\rho$ decreases as the vortex solid phase crystallizes out of the vortex liquid phase.

*3.7. H-T vortex phase diagram identifying the low-field VL, VS phases and the coexistence regime*

In different regions of the sample the upper limit of the VL phase ($H_M(T)$) is determined from the peak in the $\delta B_z$ signal measured as a function of $H$ (cf. fig. 2(b)) at different $T$. In inset (c) of fig. 5 four regions labeled 1, 2, 3 and 4 in the sample are identified. By comparing with the pinning landscape in fig. 4(d), two are located within weak pinning regions (labeled 1 and 2) and the other two are located within strong pinning (labeled 3 and 4) regions of the sample. Figure 5(a) shows the location of the field $H_M(T)$ in region 4 (the strongest pinning region). Using the criterion that at $H$ just greater than $H_{c1}$, the $\delta B_z$ begins to systematically deviate from 0 G as in fig. 2(b), we estimate the local $H_{c1}(T, r)$, where $r$ denotes the location of the region where $H_{c1}(T)$ is being determined. The crosses indicate the values of $H_{c1}(T, r)$ in region 4 of the sample. The pink line is a guide to the eye. As shown in fig. 5(a), the VL phase exists between this $H_{c1}(T)$ line and upto the $H_M(T)$ curve for region 4 (solid red line through squares) Here the cross stripped region identifies the width of the VL phase regime for the strongest pinned regions of the sample. From data recorded at different $H$ and $T$ we determine the ($H$, $T$) at which $\delta B_z$ decreases and saturates to a value = $\delta H$ (cf. at ~ 50 G in fig. 2(b)) and corresponds to the onset of a VS phase. In fig. 5(a) using the above $H$ - $T$ information we identify the boundary (open triangles with dashed line) across which VS phase sets in the regions the sample with strongest pinning (viz., region 4 as indicated in fig. 5(c)). Between this solidification line and the $H_M(T)$ line, we identify the coexistence regime where the VL and VS phases coexist for the region 4 (recall the shaded coexistence regime identified in fig. 2(b) where the $\delta B_z$ decreases from a peak value before saturating to a value = $\delta H$). As discussed earlier, a comparison of figs. 5(b) and 5(c) shows that whether the melting in a particular region of the sample is approached in an isothermal or an isofield measurement, the melting appears at a unique location in the vortex matter phase diagram viz., at ($H_M$, $T_M$). In fig. 5(d) we show DMO images captured when reversing the magnetic field from a higher field of 200 Oe at 52 K. In fig. 5(d) we observe bright contrast inside the sample in the DMO image at 45 Oe which corresponds to an enhancement encountered in the $\delta B_z$ similar to that in fig. 1. Note that the bright contrast observed due to melting across the $H_M(T)$ line observed in fig.5(d) is not as significant as in fig. 1. We believe that the diminished contrast in fig. 5(d) is a result of the $\delta B_z$ signal riding on a significantly large differential magnetization generated in the second quadrant of the hysteresis loop, viz., as a result of a sharp change in magnetization with decreasing field from 200 Oe.

Figure 5(e) shows the location of the field $H_M(T)$ in regions 1, 2, 3 and 4. From fig. 5(e) it is clear that in regions with strong pinning (3, 4) the threshold $H$ or $T$ for the transformation from a VL to VS phase is higher compared to the relatively weaker pinning regions (1, 2). The $H_M(T)$ lines for the four regions of the sample indicate that the phase boundary across which the VL phase transforms into the VS phase progressively shifts upwards with increasing pinning. It is noteworthy that at high $T$ close to $T_c$ one finds that the different $H_M(T)$ curves merge together, corresponding to the uniform appearance of the VL phase across the entire sample (see DMO images at 85 K in fig. 6). Figures 6(a) - 6(c) show DMO images of VL phase spreading across the sample at $T$ = 85 K with $H$ = 9, 12 and 15 Oe and $\delta H$ = 1Oe, respectively. Note that at 12 Oe, brightening corresponding to peak in $\delta B_z$ which appears across the entire sample and the bright contrast disappears altogether by 15 Oe. This suggests that at 85 K, the low-field first order like VL to VS transition is sharper (with a width of < 3 Oe) as compared to the broader width of the transition at 50 K, viz., a width of ~ 5 - 10 Oe. At $T$ close to $T_c$ the VL phase appears almost uniformly across the entire sample, which is due to thermal fluctuations smearing out the variations in the pinning landscape thereby making the phase transition set in more uniformly across the entire sample. The $H_M(T)$ of 12 Oe at $T$ close to $T_c$ (85 K) coincides with point on the high-field melting line in BSCCO [20,30,33], the changes in $B_z$ observed in this ($H$, $T$) regime are also comparable. In our sample there are significant effects of pinning which lead to a broadening of the transformation from the vortex liquid to solid phase at low fields. It has been shown earlier [34] that at low $T$ the jump in magnetization associated with the high field VS to VL transition is smeared out due to strong pinning induced irreversible magnetization response. This high-field jump in magnetization in this situation is discernable only after suppressing the underlying irreversible magnetization response with an in-plane oscillating magnetic field. We have made no such attempts to suppress the underlying irreversibility in magnetization to identify the high field jumps at low $T$. Therefore within the limited low-field range of our setup, at low $T$ we are unable to identify the high-field melting line in our sample which is presumably broadened out due to pinning effects. In our sample within the temperature (20 - 90 K) and field range (0 - 200 Oe) of our investigation, we have found evidence for only the low-field VL to VS phase transformation. Due to the large depinning current density at low $H$, transport measurements in this regime suffer from significant joule heating effects. Therefore it is difficult to identify the low-field VL to VS phase transition line using standard bulk transport measurement techniques as have been used for identifying the high-field melting line in the past [35]. As indicated by the dotted (blue) arrow at the edge of the cross shaded region in fig.

3(a), the VL to VS transformation may appear only as a slight bump on the bulk magnetization response making it difficult to discern the low-field VL to VS transition line from bulk magnetization measurements.

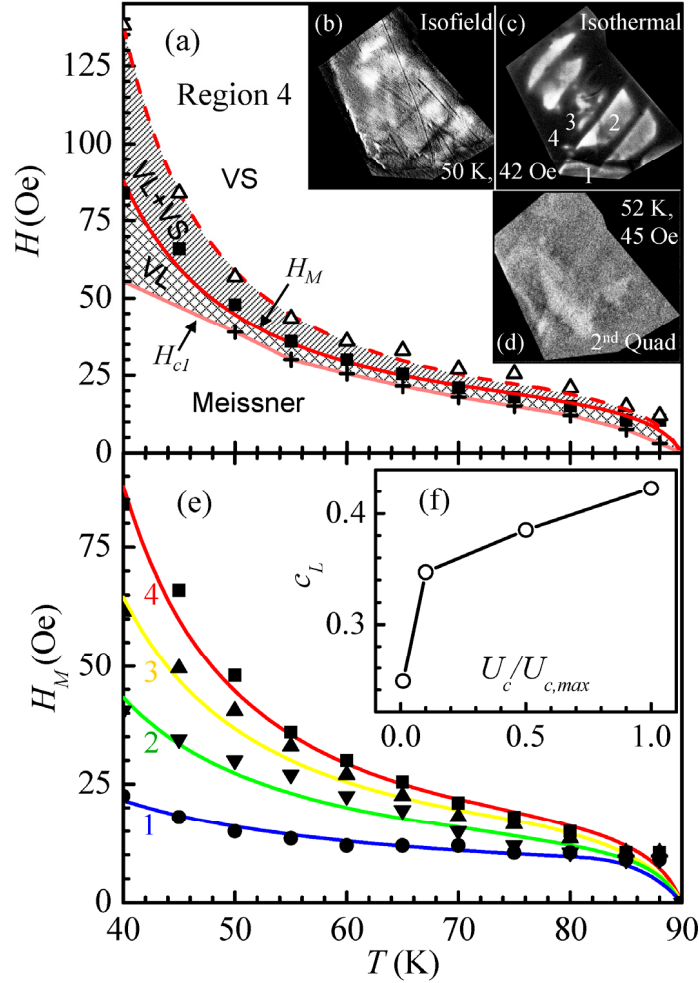

**Figure 5.** (a) Local $H$-$T$ phase diagram for the sample region 4 as marked in inset (c). Squares represent the $H_M(T)$ curve for region 4. The solid red line through the data is fit to the low-field re-entrant melting line. Crosses represent the boundary of the Meissner phase for region 4 (the pink line is a guide to the eye). The cross-striped region indicates the VL phase for region 4. The dashed (red) line represents the boundary across which VS phase appears in region 4. The cross-shaded region between the $H_M(T)$ curve and the dashed (red) boundary above which VS phase appears, indicates the VL+VS coexistence regime in region 4. Onset of the VS phase throughout the sample occurs beyond the dashed (red) line (cf. text for details). Insets (b) and (c) show DMO images at 50 K, 42 Oe. The one to the left (b) is obtained from an isofield run, and the other (c) from an isothermal run. The numbers on the image in (c) mark regions of the sample with different levels of pinning (1, 2 are weak pinning regions and 3, 4 are strong pinning regions) as identified from the $U_c$ map in fig. 4. (d) DMO image obtained at 52 K, 45 Oe, while reducing $H$ from 200 Oe, viz., from deep inside the VS phase. The patches in the DMO image with whitish contrast represent regions which have melted upon crossing $H_M(T)$ line on reversing the field form the VS phase. (e) $H_M(T)$ curves for the four different sample regions as marked in inset (c). The coloured solid lines through the data are fits to the low-field re-entrant melting line. (f) Variation of the parameter $c_L$ with normalized $U_c$.

*3.8. Fitting the theoretical expression for the low-field melting line*

Theoretically it is predicted [10] that at low fields, the boundary in the *H - T* vortex phase diagram across which the VL to VS phase transformation occurs obeys, $H_m(T) \approx \frac{\phi_0}{\lambda^2} \frac{1}{4} \left[ \ln\left( \frac{4\pi c_L^2}{(3\pi)^{1/4}} \frac{\varepsilon_0 \lambda}{T} \right) \right]^{-2}$.

Here, $\lambda$ is the penetration depth, $\varepsilon_0 = (\phi_0 / 4\pi\lambda)^2$ the vortex line energy. Usually thermal melting of the vortex state always occurs in the presence of quenched random pinning centers. The above expression is based on considering a softening of the elastic modulii of the vortex solid due to weak inter-vortex interaction at low fields which lead to enhanced wandering of flux lines from their mean position. Enhanced r.m.s. wandering of the flux lines has been considered to arise not only from thermal fluctuations but also quenched random disorder [36], leading to a net r.m.s. fluctuation of a vortex line about the mean position ($\langle u^2 \rangle_{th}^{1/2}$), which is determined phenomenologically using the Lindemann criterion viz., $c_L = \langle u^2 \rangle_{th}^{1/2} / a_0$, where $c_L$ is the Lindemann number which quantifies the typical criterion for melting, viz., when the r.m.s. fluctuation of a vortex line about its mean position becomes $c_L \sim$ 10 - 20% of the inter-vortex spacing. The $H_M(T)$ curves in figs. 5(a) and 5(e) demarcate the boundary across which the VL transforms into a VS phase via an intermediate coexistence regime, in regions of the sample with different pinning strengths. The $H_M(T)$ curves are fitted (solid coloured lines) to the above expression using $c_L$ as the only fitting parameter and using $\lambda(T) = \lambda_0 / (1 - (T/T_c)^2)^{1/2}$, with $\lambda_0 = \lambda(T = 0 \text{ K}) = 200$ nm. In fig. 5(e) we find that all the different $H_M(T)$ curves in regions of sample with different pinning strength fit to the above expression for low-field melting by considering $c_L$ as a fitting parameter. In the low-temperature regime below 85 K, the $H_M(T)$ curves are fitted with a different $c_L$. As discussed in the context of fig. 6, above 85 K, all the different $H_M(T)$ curves merge into the same main curve. This feature is denoted by the lines in fig. 5(e) representing the different fits merging into one single fitted value. The VL+VS regime and VS boundary, similar to those shown for region 4 in fig. 5(a), can be constructed for all the other regions (1, 2 and 3) of the sample (not shown for the sake of clarity in fig. 5(e)). In fig. 5(f) we plot the $c_L$ value we find from fitting of the different $H_M(T)$ lines as a function of the normalized $U_c$. We observe that the value of $c_L$ changes from close to 0.25 for low pinning regions to a value close to 0.4 for regions with strong pinning. Increase in the value of $c_L$ with increasing pinning strength indicates that in the dilute vortex regime increase in the pinning strength results in an enhanced stability of the vortex state against thermal fluctuations. The upward shift of the $H_M(T)$ line in the field temperature phase diagram with increase in pinning strength is distinct from the behaviour known for the high-field melting line, viz., increased point disorder results in shifting the high-field melting line to lower fields [37].

The changes in the configuration of vortices triggered in response to an external magnetic field modulation help us in determining the presence of a phase transformation in the vortex state. From our experiments at low fields we find a significant change in the density of vortices generated in response to a small modulation in external field which suggests a change in entropy. For the low-field VL to VS phase transition, using the Clausius-Clapeyron relation for vortex lattice melting, viz., $\Delta s = -\frac{d\phi_0}{4\pi} \frac{\Delta B_M}{B_M} \frac{dH_M}{dT}$, we estimate the change in entropy ($\Delta s$) per vortex per Copper oxide plane [8] in BSCCO $\sim 0.05 k_B$, where we use the size of the jump in *B* in the low-field VL phase, viz., $\Delta B_M = \Delta \langle \delta B_z \rangle_{bright} \sim 0.1$ G, $B_M \sim 36$ Oe at 50 K, CuO interlayer spacing (*d*) = 15 A°, a mean slope (cf. fig. 5(a)) of the low-field melting line $dH_M/dT \sim 0.8$ Oe/K. The latent heat for the low-field melting, $L \sim 0.05 k_B T_M$, appears to be comparable to that observed for high-field melting ($\sim 0.1 k_B T_M$) in BSCCO [8,9] at similar temperatures of 50 K. It appears that the pinning seems to have the opposite effect on melting of the dilute vortex state as compared to that on melting at higher fields. Point pin induced wandering of the flux lines about their mean position, especially in a soft vortex solid, promotes easier melting of the vortex solid, thereby causing the high-field melting boundary to shift down in the *H-T* vortex phase diagram with enhanced pinning [37]. Quite contrary to this behaviour known for melting at high fields [37], we observe that with enhanced pinning the low-field melting boundary moves up in the *H-T* vortex phase diagram. Therefore it appears that the vortex solid phase is shrinking both from the high-field and the low-field ends in the *H-T* phase diagram. The distinct behaviour of low-field melting line near $H_{c1}(T)$ from the high-field melting line suggests that the behavior of the two melting lines may have to treated separately theoretically.

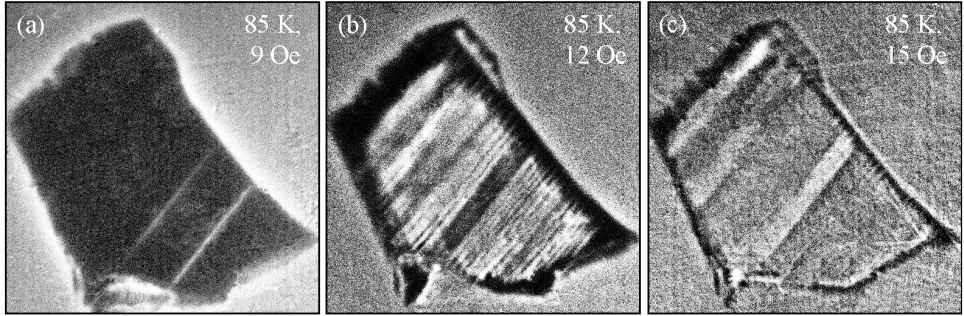

**Figure 6.** (a)-(c) DMO images at $T$ = 85 K representing uniform vortex solidification over the entire sample in a narrow field interval of 3 Oe. Solidification at 85 K is shown at (a) 9 Oe, (b) 12 Oe and (c) 15 Oe.

From figs. 1 and 6 it appears that the initial nucleation and growth of the low-field VL phase occurs along the linear defects in the sample [22] where vortex chain states [22,38] are generated by the pancake vortices in BSCCO decorating a plane of Josephson vortices which prefer to align along the linear sample inhomogeneities [31,39]. We believe that enhanced longitudinal fluctuations along the defects [40] help in thermally destabilizing the stack of pancake vortices penetrating along the Josephson vortex planes. Spread of the VL phase to sample regions with higher pinning is delayed as pinning appears to strengthen the vortex state against thermal fluctuations which is indicated by the increase in $c_L$ with pinning strength in fig. 5.

Before concluding, we would like to clarify that the anomalous brightening feature associated with a jump in $B_z$ (cf. fig. 1) is not related to an inhomogeneous vortex state. Note that regions of the sample where the anomalous jump in $B_z$ occurs (viz., regions with enhanced bright contrast in figs. 1(b), 1(c)) possess a dome shaped $B_z(r)$ field distribution (e.g., see navy blue curve corresponding to 37.5 Oe, as well as the ones above it in fig. 2(a)). We recall here that a similar dome shaped magnetic field distribution at the onset of the high-field vortex melting phenomenon in BSCCO single crystals have been reported in the past (cf. Soibel *et al*. Nature, Ref. [19]). A dome shaped $B_z(r)$ profile is characteristic of a sample with negligible bulk screening currents, with the dome shaped field distribution being associated with the screening current circulating around the sample edges (e.g., see the effect of geometrical barriers in the absence of bulk shield currents in high-$T_c$ superconductors leading to the dome shaped field profile, E. Zeldov *et al*, Ref. [23]). We argue that if an inhomogeneous vortex state were to be associated with the regions of the sample with anomalous enhancement of magneto-optical intensity, then these regions should have exhibited Bean like profiles in the local magnetic field distribution, $B_z(r)$. Consequently in these regions significant bulk screening currents (related to the gradient in Bean like $B_z(r)$ profile) should have been present along with screening currents already circulating at the sample edges. Such a current distribution would be inconsistent with observing a dome shaped field distribution. Therefore, the dome shaped field distribution we observe in fig. 2(a) suggests the absence of an inhomogeneous vortex state in the regions with the anomaly. The above assertion is also supported by our observation in fig. 4(b) inset, which shows the absence of screening currents within regions of the sample where the anomalous enhancement in magneto-optical intensity is found (cf. the dark region within the encircled region in fig. 4(b) inset). The above suggests that the observed anomalous enhancement in the magneto-optical intensity is not associated with inhomogeneity of the vortex state; rather it is a signature of a phase transformation at low vortex densities.

**4. Conclusion**

In conclusion, using magneto-optical imaging in a BSCCO single crystal, we have identified signatures of a low-field vortex liquid to vortex solid phase transition via a coexistence regime and have identified their location on the *H - T* vortex phase diagram. The first order like transition into the low-field vortex liquid from the vortex solid phase is marked by enhanced equilibrium vortex densities. The location of the low-field phases identified is found to be sensitive to the pinning strength. The locus of the low-field phase boundary in the field – temperature phase diagram fits to the theoretical curve predicted for low-field vortex liquid to solid melting. We find that the effect of enhanced pinning strength suggests that at low fields the vortex lines are stiffened against thermal wandering leading to a higher Lindemann number. We hope that the present work will stimulate future theoretical and experimental work to investigate the peculiarities of the dilute vortex phase.

**Acknowledgments**

S. S. Banerjee would like to acknowledge funding from DST, CSIR and IIT Kanpur.